\documentclass[%
prd
 ,twocolumn%
 ,secnumarabic%
,amssymb, amsmath,nobibnotes,showpacs]{revtex4}
\usepackage{bm}%
\expandafter\ifx\csname package@font\endcsname\relax\else
 \expandafter\expandafter
 \expandafter\usepackage
 \expandafter\expandafter
 \expandafter{\csname package@font\endcsname}%
\fi

\begin{document}
\title{Shortcomings of the Big Bounce derivation in Loop Quantum Cosmology}%

\author{Francesco Cianfrani$^{1}$, Giovanni Montani$^{234}$}%
\email{montani@icra.it, francesco.cianfrani@icra.it}
\affiliation{$^{1}$ICRA-International Center for Relativistic Astrophysics, Dipartimento di Fisica (G9), Universit\`a  di Roma ``Sapienza'', Piazzale Aldo Moro 5, 00185 Roma, Italy.\\
$^{2}$ Dipartimento di Fisica (G9), Universit\`a  di Roma ``Sapienza'', Piazzale Aldo Moro 5, 00185 Roma, Italy.\\  
$^{3}$ENEA C.R. Frascati (Unit\`a Fus. Mag.), Via Enrico Fermi 45, 00044 Frascati, Roma, Italy.\\
$^{4}$ICRANet C. C. Pescara, Piazzale della Repubblica, 10, 65100 Pescara, Italy.}
\date{June 2010}%

\begin{abstract}
We give a prescription to define in Loop Quantum Gravity the electric field operator related to the scale factor of an homogeneous and isotropic cosmological space-time. This procedure allows to link the fundamental theory with its cosmological implementation. In view of the conjugate relation existing between holonomies and fluxes, the edge length and the area of surfaces in the fiducial metric satisfy a duality condition. As a consequence, the area operator has a discrete spectrum also in Loop Quantum Cosmology. This feature makes the super-Hamiltonian regularization an open issue of the whole formulation. 

\end{abstract}


\maketitle

\section{Introduction} 
Loop Quantum Gravity (LQG)\cite{revloop} constitutes the most compelling attempt toward a complete non-perturbative quantum theory for the gravitational field. The key features at the ground of this scheme are both the emergence of a local SU(2) gauge invariance at the Hamiltonian level \cite{Ash,prl} and the quantization of the corresponding holonomy-flux algebra \cite{ALMMT95}. The most relevant issue is the prediction of discrete spectra for geometrical operators at the kinematical level \cite{discr}.  However, a proper implementation of the dynamics together with the characterization of semi-classical states has not been obtained yet. A path-integral formulation via spin-foam models looks promising, although several unsolved issues remain \cite{rovrec}. 

The difficulties of the general theory for gravity can be overwhelmed in the minisuperspace models, where some degrees of freedom are frozen out. In particular, the quantum description of a homogeneous and isotropic cosmological space-time has the advantage that only one variable, the scale factor $a$, parametrizes the configuration space. At the same time, this symmetric case can be regarded as an outstanding scenario because it aims to describe the early Universe dynamics at least as a first approximation (indeed there is no indication that in the quantum phase the Universe must be close to a isotropic and homogeneous configuration \cite{gp}). In this respect, an answer can be given to the most important issue that the Friedman-Robertson-Walker (FRW) dynamics leaves unsolved at a quantum level \cite{qc}, the nature of the initial singularity.  

The first cosmological application of LQG was developed in terms of invariant connections \cite{lqc}, {\it i.e.} a restriction was made to connections which respected the global homogeneity and isotropy, and this model was denoted by Loop Quantum Cosmology (LQC). Within this scheme, it has been demonstrated that the inverse scale factor was bounded from above on the zero volume eigenstates and that the super-Hamiltonian constraint became a non-singular difference equation. This features stand as good indications that the singularity is removed. Indeed, the above mentioned property of the inverse scale factor does not hold in LQG \cite{thbr} and this makes the relationship between the fundamental and the minisuperspace theory a tantalizing subject of investigation (see also \cite{engle}).

Then, in \cite{ash} a complete dynamical picture is realized by restricting to holonomies along straight lines in the fiducial metric, and so reducing the Hilbert space to the one of quasi-periodic functions. The regularization of the super-Hamiltonian takes place by fixing a fundamental length for the graphs on which the super-Hamiltonian is evaluated. Hence, the specific value of such a length is inferred from requiring that the minimum area on which the field strength of SU(2) connections is regularized coincides with the minimum area eigen-value of LQG \cite{ash1}.   

The main achievements of this procedure \cite{rob} are the avoidance of the cosmological singularity and the prediction of a bounce occurring at a certain value of matter energy density, the so-called critical density (see \cite{phen} for a phenomenological description).

As outlined in \cite{haro}, the regularization itself produces the bounce, rather than the quantization procedure. The justification of such a regularization via the requirement of a minimum area spectrum moves LQC away from LQG, where the discretization occurs already at a kinematical level, while the regularization is intimately connected with the definition of the super-Hamiltonian in the Hilbert space \cite{qsd} (a similar criticism is made in \cite{djmp}, while for a different objection based on the investigation of inverse volume corrections see \cite{clqc}). 

In this work, we elucidate the relationship between LQC and LQG, by demonstrating that a proper operator corresponding to $p$, where $|p|=a^2$, can be defined for holonomies along straight lines. The consistency between such an operator and the symplectic structure in the minisuperspace leads to fix a fundamental duality between the edge length on which holonomies are evaluated and the area of surfaces across which fluxes are defined. Furthermore, the discretization of the geometrical operators is a direct consequence of the compactness of the gauge group and it has no relation at all with the existence of a fundamental edge length. This feature prevents to follow the regularization procedure of the super-Hamiltonian adopted in \cite{ash}, \cite{ash1}.  

Henceforth, we define a map from the reduced holonomy-flux algebra to the one proper of quasi-periodic functions via the trace operator. This step concludes the derivation of the kinematics of LQC from that of LQG restricted to the FRW-like connections.   

Finally, it is outlined how in this scenario it is possible to relate the parameter at which the regularization of the super-Hamiltonian occurs with the total number of vertices of the fundamental graph underlying the classical description of the cosmological space-time. 
  
%The manuscript is organized as follows: in Sec. \ref{1}, we review the LQC formulation, focusing our attention on the definition of the Hilbert space and on the regularization procedure. Then, in Sec. \ref{2}, the $p$ operator is defined in the Hilbert space of connections along straight lines in the fiducial metric and a proper correspondence between length of edges and area of surfaces is demonstrated. Furthermore, the discreteness of the area spectrum and the independence from the length of edges is emphasized. The map from the adopted Hilbert space to the one of LQC is defined in Sec. \ref{3}. Concluding remarks follow in Sec. \ref{4}.

\section{Loop Quantum Cosmology.}\label{1}

A cosmological space-time is assumed to be homogeneous and isotropic. The metric which is compatible with these assumptions is the Friedman-Robertson-Walker one, {\it i.e.}  
\begin{equation}
ds^2=-dt^2+a(t)^2\left(\frac{1}{1+kr}dr^2+r^2d\theta^2+r^2\sin^2\theta d\phi^2\right),\label{FRW}
\end{equation}

where $k=1,0,-1$ for a closed, flat and open Universe, respectively. It is worth noting that the scale factor $a$ is the only dynamical variable, which on spatial hypersurfaces behaves as a conformal factor for the fiducial line element
\begin{equation}
{}^0\!dl^2=\frac{1}{1+kr}dr^2+r^2d\theta^2+r^2\sin^2\theta d\phi^2.
\end{equation}

LQC is based on fixing Ashtekar-Barbero-Immirzi connections and densitized 3-bein vectors as follows 
\begin{equation}
A^a_i=c{}^0\!e^a_i,\qquad E^i_a=p\sqrt{{}^0\!h}{}^0\!e^i_a,\label{AE}
\end{equation}

where ${}^0\!e^a_i$ and ${}^0\!e_a^i$ denote 3-bein vectors of the fiducial metric ${}^0\!h_{ij}$ and their inverses, respectively, while
\begin{equation}
|p|=a^2,\qquad c=\frac{1}{2}(k+\gamma \dot{a}).
\end{equation}

Within this scheme, $c$ and $p$ are fundamental phase variables and the Poisson brackets between each other are as follows (we work in units $\hbar=c=1$)
\begin{equation}
\{c,p\}=\frac{8\pi G\gamma}{3V_0},\label{pb}
\end{equation}

$V_0$ being the volume of the fiducial metric. Usually a rescaling $c\rightarrow V_0^{1/3}c$, $p\rightarrow V_0^{2/3} p$ is performed, such that $V_0$ does not appear into Poisson brackets. Here, we will not consider such a rescaling. 

The quantization is based on choosing almost periodic functions $N_\mu=e^{\frac{i\mu c}{2}}$ as a basis in the configuration space. The algebra generated by $\{N_\mu,p\}$ plays the role of the holonomy-flux algebra of LQG, such that by the analogous construction of the general case the Hilbert space turns out to be $\textsc{H}=L^2(\textbf{R}_{Bohr},d\mu_{Bohr})$, $\textbf{R}_{Bohr}$ being the Bohr compactification of the real line. In such a Hilbert space the measure is given by 
\begin{equation}
<N_{\mu'}|N_\mu>=\delta_{\mu',\mu},\label{dk}
\end{equation}

and the action of fundamental operators reads
\begin{equation}
\hat{N}_\mu\psi(c)=e^{\frac{i\mu c}{2}}\psi(c),\qquad \hat{p}\psi(c)=-i\frac{8\pi\gamma l_P^2}{3}\frac{d}{dc}\psi(c).
\end{equation}

$l_P$ being the Planck length. The expression of the super-Hamiltonian in a proper factor ordering is given by
\begin{eqnarray} 
\mathcal{H}^{\bar{\mu}}=-\frac{3V_0}{8\pi \gamma l_P^2\bar{\mu}^2}\hat{p}^{1/2}\hat\sin^2{\bar{\mu}c},\label{suph}
\end{eqnarray}

where the parameter $\bar\mu$ is non-vanishing (the limit of $\mathcal{H}^{\bar{\mu}}$ as $\bar\mu$ goes to $0$ does not exists) and this feature is taken as a reminder of the fundamental discrete structure proper of LQG. In fact, a possible way to fix $\bar\mu$ consists of assuming that the corresponding area operator, which is given by
\begin{equation}
A(\bar\mu^2)N_{\bar{\mu}}=|p|\bar\mu^2N_{\bar\mu},%%verifica
\end{equation}

reproduces the minimum eigen-value of the same operator in LQG \cite{discr}, so having \cite{ash1}
\begin{equation}
\bar{\mu}^2|p|=2\sqrt{3}\pi\gamma l_P^2.%%%%verifica
\end{equation}

This choice is particularly useful, since a consistent cosmological dynamics with a bounce replacing the initial singularity is predicted when a clock-like scalar field is introduced. Indeed, the following alternative prescription is present in literature \cite{ash}:
\begin{equation}
\bar{\mu}^2=2\sqrt{3}\pi\gamma l_P^2.
\end{equation}

This proposal was discarded, because in this case the critical density depends on the momentum of the clock-like scalar field.

\section{Phase-space variables}\label{2}

The most general connections and momenta compatible with the FRW metric (\ref{FRW}) are obtained from the expressions (\ref{AE}) by a generic SU(2) transformation. This means that although the metric has been partially fixed, nevertheless the local SU(2) gauge symmetry is not lost (this is not surprising, because such gauge transformations are related with rotations in the tangent space).
  
Let us now depict a possible description of a cosmological space-time in terms of LQG variables. 
Holonomies $h^a_\alpha$ are now being evaluated along straight edges $\alpha$ parallel to ${}^0\!e^i_a$, so finding 
\begin{equation}
h^a_\alpha=e^{i\mu c {}^j\!\tau_a},\label{hol}
\end{equation}

$\mu$ being the edge length, $\mu=\int_\alpha {}^0\!e^a_i\frac{d\alpha^i}{dt}dt$, while ${}^j\!\tau_a$ denotes the SU(2) generator in the $j$-representation. In what follows we will label the holonomies by $h^a_\mu$.  

Similarly, fluxes $E_a(S)$ are restricted to those ones across surfaces $S$, $x^i=x^i(u,v)$, whose normal coincides with ${}^0\!e^a_i$ and their classical expression reads  
\begin{equation}
E_a(S)=p\Delta,\qquad \Delta=\int_S {}^0\!e^i_a\epsilon_{ijk}\partial_ux^j\partial_vx^kdudv,
\end{equation}

where $\Delta$ gives the flux of ${}^0\!e^i_a$ through $S$. $\Delta$ measures the area of $S$ itself in the fiducial metric and in the following it will be used as a label for $E_a$.

If $S$ and $\alpha$ intersects each other, the flux operators act on holonomies as follows
\begin{equation}
\hat{E}_a(\Delta)h^b_\mu=8\pi\gamma l_P^2h^b_\mu {}^j\!\tau_a \delta^a_b\textrm{sign}{\Delta\mu}\label{E} 
\end{equation}

where in the last relation repeated indexes are not summed.

Substituting the expression for $E_a(S)$ in terms of $p$, one finds 
\begin{equation}
\hat{p}\Delta h^a_\mu=8\pi\gamma l_P^2h^a_\mu {}^j\!\tau_a\textrm{sign}{\Delta\mu},\label{pdelta}
\end{equation}

but from the Poisson bracket (\ref{pb}) the operator $p$ can be represented in the form
\begin{equation}
\hat{p}=-i\frac{8\pi\gamma l_P^2}{3V_0}\frac{d}{dc},
\end{equation}  

whose action on holonomies (\ref{hol}) gives
\begin{equation}
\hat{p}h^a_\mu=\frac{8\pi\gamma l_P^2\mu}{3V_0}h^a_\mu{}^j\!\tau_a.\label{p} 
\end{equation}

Therefore, relations (\ref{pdelta}) and (\ref{p}) are consistent when
\begin{equation}
|\Delta\mu|=3V_0.
\end{equation}

\emph{This relation fixes a fundamental duality between the length of the edges along which holonomies are evaluated and the area of the surfaces across which fluxes are defined.}

\section{Quasi-periodic functions}\label{3}

Within this scheme it is possible to establish a clear correspondence between the Hilbert space generated by holonomies (\ref{hol}) and the one of quasi-periodic functions. This correspondence can be realized via the trace on SU(2) indexes. 

In fact, tracing both sides of Eq. (\ref{E}) %and of the same expression times. This way 
one gets
\begin{eqnarray} 
tr(E_a(S)h^a_\mu)=2\hat{p}|\Delta|\Sigma_{n=0}^{j-\theta}\cos{(\mu c(n+\theta))}=\nonumber\\=8\pi\gamma l_P^2 tr(h^a_\mu {}^j\!\tau_a)=-16\pi\gamma l_P^2\Sigma_{n=0}^{j-\theta}n\theta\sin{(\mu c(n+\theta))},
%\\tr(E_a(S)h^a_\mu{}^j\!\tau_a)=p\Delta\Sigma_{n=0}^{j/\alpha}n\sin{(\mu cn\alpha)}=\nonumber\\=tr(h^a_\mu {}^j\!\tau_a{}^j\!\tau_a)=\Sigma_{n=0}^{j/\alpha}n^2\cos{(\mu cn\alpha)},
\end{eqnarray}

where $\theta=1/2,0$ for $j$ half-integer and integer, respectively.

It is worth noting that after the trace has been performed, linear combinations of quasi-periodic functions come out.
 
The action of $\hat{p}$ on such quasi-periodic functions reads as
\begin{equation} 
\hat{p}e^{i\widetilde{\mu}c}=\frac{8\pi \gamma l_p^2}{3V_0}\widetilde\mu e^{i\widetilde{\mu}c}.\label{pmu}
\end{equation}

In LQG two kind of information are present, the one related with the edge length $\mu$ and the one giving the SU(2) quantum number $n$. These two notions are condensed in the factor $\widetilde\mu=n\mu$, such that the SU(2) gauge structure is not manifest. However such an information is required to infer the area spectrum.

%The scalar product between holonomies (\ref{hol}) based on a fixed point gives exactly the Kroenecker delta, as for the Bohr compactification space (\ref{dk}).
In fact, within this scheme, the regularized area operator can be represented by the square root of $\hat{p}^2\Delta^2$, thus its action on quasi periodic functions is 
\begin{equation}
\hat{A}e^{i\mu nc}=\sqrt{\hat{p}^2\Delta^2}e^{i\mu nc}=8\pi \gamma l_p^2\theta |n|e^{i\mu nc}.
\end{equation}

Hence, the area operator has a discrete spectrum whatever value takes the parameter $\mu$. The spectrum does not coincide with the one of the fundamental theory \cite{discr}, which is related with the Casimir of the SU(2) group. 

Therefore, the procedure adopted in \cite{ash} to infer the parameter $\bar\mu$ required for the super-Hamiltonian regularization cannot be justified on the level of the area discrete spectrum. By other words, the existence of a minimum value for $\mu$ is not a consequence of fundamental properties of LQG and this short-coming of the previous derivation leaves open the question about the proper implementation of the dynamical constraint.

The regularized super-Hamiltonian takes the following expression in LQG \cite{qsd}
\begin{eqnarray}
H=-\frac{1}{32\pi^2\gamma^3 l_P^4}\sum_v H_v,\\ H_v=-\epsilon^{ijk}Tr[h(s_{ij
})h(s_k)[V,h^{-1}(s_k)]],\label{sH}
\end{eqnarray}

where the sum is on all vertices $v$ of the graph on which $H$ acts. Here 
$s_{ij}$ denotes the square emerging from $v$ with the edges along the directions $ij$, while $s_k$ is the edge along the direction $k$. All holonomies in the expression (\ref{sH}) are in the fundamental representation. $V$ is the volume operator in the full space.

The restriction to a FRW space-time implies to replace $V$ and $h(s_i)$ with $\hat{p}^{3/2}V_0$ and $h_{\bar\mu}^a$, respectively, $\bar\mu$ being the value at which the regularization should take place. From Eq. (\ref{p}) one finds 
\begin{equation}
[V,h_{\bar\mu}^a]=V_0[\hat{p}^{3/2},h_{\bar\mu}^a]=8\pi\gamma\bar\mu l_P^2 \hat{p}^{1/2}{}^{1/2}\!\tau_a h_{\bar\mu}^a,
\end{equation} 

which reproduces the following expression when inserted into the super-Hamiltonian (\ref{sH})
\begin{eqnarray}
H=-\sum_v \frac{3\bar\mu}{8\pi l_P^2 \gamma^2}\hat{p}^{1/2}\hat\sin^2{\bar{\mu}c}.
\end{eqnarray}

If we assume that each vertex gives the same contribution, then $H$ can be written as
\begin{eqnarray}
H=-\frac{3N_v\bar\mu^3}{8\pi l_P^2\gamma^2\bar\mu^2}\hat{p}^{1/2}\hat\sin^2{\bar{\mu}c},\label{shv}
\end{eqnarray} 

$N_v$ being the total number of vertices of the fundamental graph underlying the continuous space-time manifold. It is worth noting that the two expressions (\ref{suph}) and (\ref{shv}) coincide if
\begin{equation}
V_0=N_v\bar\mu^3\rightarrow \bar\mu=\left(\frac{V_0}{N_v}\right)^{1/3}.
\end{equation} 

Therefore, the assumption that the regularized super-Hamiltonian retains the same expression as in \cite{ash} links $\bar\mu$ with the total number of vertices.

\section{Conclusions}\label{4}

We analyzed the possibility to infer LQC from the general framework of LQG. In particular, we outlined that the proper global operators could be defined as soon as the restriction to FRW-like connections and momenta took place. However, a fundamental condition linked the area of the surfaces across which fluxes were defined and the length of the edges along which holonomies were evaluated. Such a relation allowed to avoid the presence of the parameter $\mu$ in the spectra of geometrical operators, so reconciling LQC with the local character proper of the LQG formulation. Moreover, we pointed out that by tracing on SU(2) indexes the Hilbert space of quasi-periodic functions came out.  

Therefore, the findings of this work exclude the possibility to connect the regularization procedure of the super-Hamiltonian with the kinematical properties of the full theory. 

Furthermore, the adopted procedure allowed us to infer the super-Hamiltonian constraint from the properties of the graph underlying the classical continuous description of the space-time manifold. In particular, a fundamental connection has been established between the parameter $\bar\mu$ at which the regularization took place and the total number of vertices. This feature confirms the point of view adopted in \cite{clqc} that the regularization of the super-Hamiltonian is deeply connected with full LQG such that $\bar\mu$ is an ambiguity in LQC.     

However a different approach to define a consistent LQC is described in \cite{clqc}, where a local definition of cosmological quantities is suggested via the introduction of local patches. Within this scheme, in each local patch the duality between the area of the surfaces and the edge length would be still realized, but actually $|\Delta\mu|=V_P$, $V_P$ being the patch volume in the fiducial metric. 
%Furthermore building blocks of any space description via spin-networks are patches containing a single node, whose volume is given by the cube of the edge length in the fiducial metric, together with fluxes and momenta defined in each patch. This scenario leads to \begin{equation}\Delta<a^.   \end{equation}

\end{document}